\documentclass[reqno,12pt,a4paper]{amsart}
\usepackage{tikz}
\usepackage{tikz-3dplot}
\usepackage{marvosym}
\usepackage{euscript}
\usepackage{multicol}
\usepackage{amsmath}
\usepackage{times}

\usetikzlibrary{shapes.geometric}

\usepackage{amssymb}

\usepackage{euscript}
\usepackage{multicol}
\usepackage{amsmath}
\usepackage{times}

\date{}

\begin{document}

\title[Lyapunov exponent  for   quantum graphs]{{\vspace*{-1.5cm}\sc Lyapunov exponent  for     quantum graphs  that are elements of  a  subshift of finite type}}

\author{Oleg Safronov}
\email{osafrono@charlotte.edu}
\address{Department of Mathematics and Statistics,  UNCC,   Charlotte,  NC}

\maketitle

\begin{abstract}We consider  the Schr\"odinger operator on the  quantum graph  whose edges  connect  the points of ${\Bbb Z}$. The numbers of the  edges  connecting two  consecutive  points $n$ and $n+1$
are   read along the orbits of a shift  of  finite  type. 
We   prove that  the Lyapunov exponent is  potitive  for  energies $E$ that do not belong to a  discrete  subset  of $[0,\infty)$.
The number of  points $E$ of this  subset in $[(\pi (j-1))^2, (\pi j)^2]$  is the same  for all $j\in {\Bbb N}$.
\end{abstract}






\newtheorem{theorem}{Theorem}
\newtheorem{lemma}[theorem]{Lemma}
\newtheorem{proposition}[theorem]{Proposition}
\newtheorem{example}[theorem]{Example}
\newtheorem{definition}[theorem]{Definition}
\newtheorem{corollary}[theorem]{Corollary}
\newtheorem{remark}[theorem]{Remark}
\newtheorem{assumption}[theorem]{Assumption}
\newtheorem{conjecture}[theorem]{Conjecture}
\newtheorem{claim}[theorem]{Claim}

\renewcommand{\thefootnote}{\ensuremath{\fnsymbol{footnote}}}

\def\theequation{\arabic{section}.\arabic{equation}}

\def\define{\stackrel{\Delta}{=}}
\def\real{\mathbb{R}}
\def\Real{\mathbb{R}}
\def\R{\mathbb{R}}
\def\endproof{\hfill\diamond}
\def\sB{{\cal B}}
\def\sF{{\cal F}}
\def\sG{{\cal G}}
\def\sf{{\cal F}}
\def\sg{{\cal G}}
\def\sH{{\cal H}}
\def\sL{{\cal L}}
\def\sh{{\cal H}}
\def\sK{{\cal K}}
\def\sC{{\cal C}}
\def\sQ{{\cal Q}}
\def\Prob{\mathbb{P}}
\def\Prob{\mathbb{R}}
\def\Expn{\mathbb{E}}
\def\P{\mathbb{P}}
\def\E{\mathbb{E}}
\def\Ind{I}
\def\ind{I}
\def\Wtilde{\widetilde{W}}
\def\Btilde{\widetilde{B}}
\def\Ptilde{\widetilde{\Prob}}
\def\Etilde{\widetilde{\Expn}}
\def\Ltilde{\widetilde{L}}
\def\ptilde{\tilde{p}}
\def\qtilde{\tilde{q}}
\def\Ntilde{\widetilde{N}}
\def\ntilde{\tilde{n}}
\def\btilde{\tilde{b}}
\def\sgn{\mbox{\rm sgn}}

\bigskip

\section{Statement of  the main results of the paper  -- Theorems~\ref{thm2}, \ref{thm3} and  ~\ref{thm5}}

For a  positive integer  $\ell>1$,
let  $\Omega$  be   the compact metric  space whose elements are  infinite sequences $\{\omega_n\}_{n\in {\Bbb Z}}$  such that $\omega_n$ is an integer between $1$ and $\ell$.
Put differently,  $\omega_n\in \{1,\dots,\ell\}={\mathcal A}$ for each $n$.
To make it  more complicated, we  will assume   that there  are sequences in ${\mathcal A}^{\Bbb Z}$  that are not allowed to be in $\Omega$ and we assume that forbidden words are of length $2$.
The   metric $d(\cdot,\cdot)$  on $\Omega$  is  defined by  $$d(\omega,\omega')=e^{-N(\omega,\omega')},$$
where $N(\omega,\omega')$ is  the  largest  nonnegative integer   such that $\omega_n=\omega_n'$   for all $|n|< N(\omega,\omega')$.
Define the mapping      $T:\,\Omega\to\Omega$     by
$$
\bigl(T\omega\bigr)_n=\omega_{n+1},\qquad  \forall n\in {\Bbb Z}.
$$
Such a mapping $T$ is called a subshift of finite type.

For each $\omega\in \Omega$, we will construct  the  graph $\Gamma_\omega$,
 displayed  below for the  case  where $\ell=2$  and $\omega=\dots2,1,1,2,2,2,1,1,2,1,2,1,1,1,\dots$

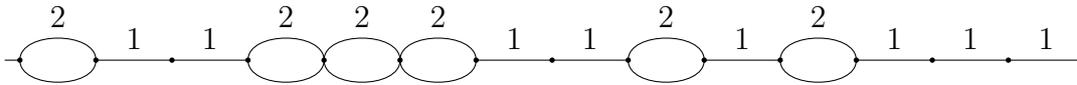
\begin{tikzpicture}
\fill[black] (-3,0) circle (1pt) (-2,0) circle (1pt) (-1,0) circle (1pt) (0,0) circle (1pt) (1,0) circle (1pt) (2,0) circle (1pt)  (3,0) circle (1pt)   (4,0) circle (1pt) (5,0) circle (1pt) (6,0) circle (1pt) (7,0) circle (1pt) (8,0) circle (1pt) 
(9,0) circle (1pt)  (10,0) circle (1pt);
\draw (-3.2,0) --  (-3,0) ;
\draw (-3,0)to[out=90,in=90] node[midway,above] {$2$} (-2,0);
\draw (-3,0)to[out=-90,in=-90]  (-2,0);

\draw (-2,0) -- node[midway,above] {$1$}  (-1,0) ;

\draw (-1,0) --  (0,0) node[midway,above] {$1$} ;

\draw (0,0)to[out=90,in=90] node[midway,above] {$2$} (1,0);
\draw (0,0)to[out=-90,in=-90]  (1,0);
\draw (1,0)to[out=90,in=90] node[midway,above] {$2$} (2,0);
\draw (1,0)to[out=-90,in=-90]  (2,0);
\draw (2,0)to[out=90,in=90] node[midway,above] {$2$} (3,0);
\draw (2,0)to[out=-90,in=-90]  (3,0);
\draw (3,0) -- node[midway,above] {$1$}  (4,0) ;

\draw (4,0) --  (5,0) node[midway,above] {$1$} ;
\draw (5,0)to[out=90,in=90] node[midway,above] {$2$} (6,0);
\draw (5,0)to[out=-90,in=-90]  (6,0);
\draw (6,0) -- node[midway,above] {$1$}  (7,0) ;
\draw (7,0)to[out=90,in=90] node[midway,above] {$2$} (8,0);
\draw (7,0)to[out=-90,in=-90]  (8,0);
\draw (8,0) -- node[midway,above] {$1$}  (9,0) ;
\draw (9,0) -- node[midway,above] {$1$}  (10,0) ;
\draw (10,0) -- node[midway,above] {$1$}  (11,0) ;

\end{tikzpicture}

Namely,
let ${\Bbb Z}$ be the  set of integer numbers.  For each  $\omega\in \Omega$ and  $n\in {\Bbb Z}$,  we  consider  $\omega_n$ copies  of the   interval $[n,n+1]$.
Denoting    these  copies  by $I_{n,j}$, where $j=1,\dots,\omega_n$,
we define the graph $\Gamma_\omega$   as the union
\[
\Gamma_\omega=\bigcup_{n\in {\Bbb Z}}\Bigl(\bigcup_{j=1}^{\omega_n}I_{n,j}\Bigr).
\]
While  the interiors of  the intervals $I_{n,j}$ are assumed to be disjoint,  we will also assume that
their   endponts are shared in the  sense  that $n$ is the common left endpoint,  and $n+1$ is   the common right  endpoint of the intervals  $I_{n,j}$.   Thus,
\[
\bigcap_{j=1}^{\omega_n}I_{n,j}=\{n\}\cup \{n+1\}.
\]

There is a natural Lebesgue measure on $\Gamma_\omega$  whose  restriction to each  $I_{n,j}$ is the Lebesgue measure on  this interval.
The main   object  of our  study  is   the Schr\"odinger  operator $H_\omega$  formally  defined  by
 \[H_\omega u =-u''\]  on the domain $D(H_\omega)$  consisting of  certain absolutely continuous functions on the  graph $\Gamma_\omega$.  
 Namely,  let $u_{n,j}$  be the restriction of $u$  to the interval $I_{n,j}$.
Then 
\[
u\in D(H_\omega),
\]
if an  only if,  all functions  $u_{n,j}$   belong to the Sobolev spaces $W^{2,2}(I_{n,j})$,  the    functions  $u$ and $u''$ are square integrable,   and for each $n\in {\Bbb Z}$,
the sum  of  derivatives of $u$ in all outgoing from  $n$ directions is zero:
\begin{equation}\label{Kirchhoff}
\sum_{j=1}^{\omega_n}u'_{n,j}(n) =\sum_{j=1}^{\omega_{n-1}}u'_{n-1,j}(n), \quad \forall n\in {\Bbb Z}.
\end{equation}
The  last  condition is  called Kirchhoff's   condition  at the point $n$.  Note  that the operator $H_\omega$ is self-adjoint in the space $L^2(\Gamma_\omega)$.

\begin{proposition} Let $k> 0$  be different  from integer multiples of $\pi$.
Let $\phi$  be the solution of the equation
\[
-\phi''=k^2 \phi,\qquad \text{on}\quad [0,1].
\]
Then
\[
\phi(x)= \frac1{\sin k}\bigl(\sin(k(1-x))\phi(0)+ \sin(kx)\phi(1)\bigr).
\]
In particular,
\[
\phi'(0)= \frac{k}{\sin k}\bigl(-\cos(k)\phi(0)+ \phi(1)\bigr)\qquad \text{and}\qquad \phi'(1)= \frac{k}{\sin k}\bigl(-\phi(0)+ \cos(k)\phi(1)\bigr).
\]

\end{proposition}

\begin{corollary}  Let $k\neq \pi j$  for any $j\in {\Bbb Z}$.  Let $u$   be the absolutely  continuous  solution of the  equation
\[
-u''(x)=k^2 u(x),\qquad  \text{for  a.e. } \quad  x\in \Gamma_\omega.
\]
satisfying  Kirchhoff's condition \eqref{Kirchhoff}. Then
\begin{equation}\label{SEquation}
\omega_n u(n+1)+\omega_{n-1}u(n-1) -(\omega_n+\omega_{n-1}) \cos(k) u(n)=0.
\end{equation} 
\end{corollary}

Spectral properties of $H_\omega$ are related to the 
behavior of  solutions to the equation
\eqref{SEquation}.
On the other hand,   all  solutions
to \eqref{SEquation} can be described in terms of the cocycles  $(T, A)$  with  $A=A^{(k)}:\, \Omega\to{\rm SL}(2, {\Bbb R})$ defined by
\begin{equation}\label{defineA}
 A(\omega)=
 A^{(k)}(\omega)= \sqrt{\frac{\omega_{0}}{\omega_{-1}}} \begin{pmatrix}
\frac{\omega_0+\omega_{-1}}{\omega_0}\cos(k)& -\frac{\omega_{-1}}{\omega_0}\\
1& 0
\end{pmatrix}
\end{equation}
Namely,  $u$ is a solution of \eqref{SEquation} if  and only if
$$
\begin{pmatrix} u(n)\\ u(n-1) \end{pmatrix}= \frac{\omega_{-1}}{\omega_{n-1}}A_n(\omega)\cdot  \begin{pmatrix} u(0)\\ u(-1) \end{pmatrix},\qquad \forall n\in {\Bbb Z},
$$
where
$$
A_n(\omega)=\begin{cases}A(T^{n-1}\omega)\cdots A(\omega)\quad \text{if}\quad n\geq 1;\\
[A_{-n}(T^n\omega)]^{-1}
\quad \text{if}\quad n\leq -1;\\ 
{\rm Id}\quad \text{if}\quad n=0.
\end{cases}
$$

{\it Definition}.  A  function $A:\, \Omega\to {\rm SL}(2, {\Bbb R})$  is said to be locally constant,   if there is an $\epsilon>0$  such that
$$
A(\omega')=A(\omega)\qquad  \text{whenever}\qquad d(\omega',\omega)<\epsilon.
$$

\bigskip

Clearly, $A$ defined  by \eqref{defineA}  is locally constant.

Since  $\Omega$  is a metric space,   we  can talk about 
 the Borel $\sigma$-algebra of  subsets of $\Omega$  and consider probability measures on $\Omega$.
Let $\mu$  be a $T$-ergodic  probability   measure on $\Omega$. The Lyapunov exponent  for $A$ and $\mu$
is defined by
$$
L(A,\mu)=\lim_{n\to\infty}\frac1n \int \ln(\|A_n(\omega)\|)d\mu(\omega).
$$
Clearly,  $L(A,\mu)\geq 0$.
By Kingman’s subaddive ergodic theorem, 
$$
\frac1n \ln(\|A_n(\omega)\|)\qquad \text{converges to}\quad L(A,\mu)\qquad \text{as}\quad n\to \infty,
$$
for $\mu$-almost every $\omega\in \Omega$.  For  simplisity,  we  write $L(k)=L(A,\mu)$.

Our  main theorem   gives  sufficient   conditions guaranteeing   that  the set
\begin{equation} \label{DefL} {\frak L}(A,\mu)=\bigl\{k\in [0,\pi]:\,\, L(A,\mu)=0\bigr\}
\end{equation}
is  finite.  One of these conditions is   that $\mu$ has  a local product structure.

Let us now give a  formal  definition of  a measure  having  this  property.
We first define  the   spaces  of semi-infinite  sequences
$$
\Omega_+=\{\{\omega_n\}_{n\geq0}:\,\, \omega\in \Omega\}
\quad\text{
and 
}\quad
\Omega_-=\{\{\omega_n\}_{n\leq0}:\,\, \omega\in \Omega\}.
$$
Then using  the natural projection  $\pi_\pm$
 from $\Omega$ onto $\Omega_\pm$,  we  define  $\mu_\pm=(\pi_\pm)_*\mu$  on $\Omega_\pm$  to
 be
the pushforward measures of $\mu$.  After  that,  for each $1\leq j\leq \ell$, we introduce
 the
cylinder sets
$$
[0;j]=\{\omega\in\Omega:\,\, \omega_0=j\}\quad
\text{and}\quad
[0;j]_\pm=\{\omega\in\Omega_\pm:\,\, \omega_0=j\}.
$$ 
A local product structure  is  a relation between   the measures $\mu_j=\mu\bigl|_{[0;j]}$  and   the measures $\mu_j^\pm=\mu_\pm\bigl|_{[0;j]}$.
To describe this relation, we need  to consider
the 
natural homeomorphisms
$$P_j: [0;j]\to [0;j]_-\times [0;j]_+   $$ 
defined  by  $$P_j(\omega)=\bigl(\pi_-\omega,\pi_+\omega\bigr),\qquad \forall \omega\in \Omega.$$

\smallskip

{\it Definition}. We say that $\mu$  has a local product structure if there is a positive   $\psi:\Omega\to (0,\infty)$
such that for each $1\leq j\leq \ell$, the function  $\psi\circ P_j^{-1}$  belongs to $ L^1\bigl( [0;j]_-\times [0;j]_+, \mu^-_j\times\mu^+_j\bigr)$
and
$$
\bigl(P_j\bigr)_*d\mu_j=\psi\circ P_j^{-1}\, d( \mu^-_j\times\mu^+_j).
$$

\begin{theorem}\label{thm2}
Suppose $T:\Omega\to\Omega$  is a subshift of finite type  and $\mu$ is a $T$-ergodic
probability measure that has a local product structure and posseses the property ${\rm supp }\,\mu=\Omega$.   Suppose $T$ has a fixed point, and at least one $\omega\in \Omega$ that is  not a fixed point.   Then the set ${\frak L}(A,\mu)$ is finite.
\end{theorem}

Theorem~\ref{thm2} could be  viewed as an analogue of  Theorem 1.2 of the paper \cite{ADZ} where  the authors  consider
the discrete Schr\"odinger operator with a real potential $n\mapsto V(T^n\omega)$  on ${\Bbb Z}$.
The function  $V:\Omega\to {\Bbb R}$  is assumed to be locally constant.

In the   theorem below,  we assume that the length of all forbidden words is  two.  A  word is  said to be admissible provided it is   present  at least in one $\omega\in \Omega$.

\begin{theorem}\label{thm3} Let  $T:\Omega\to\Omega$  be  a subshift of finite type  and $\mu$ be  a $T$-ergodic
probability measure that has a local product structure and posseses the property ${\rm supp }\,\mu=\Omega$.   Suppose that  there are two distinct letters $j_0,j_1$ in ${\mathcal A}$ such that the words   $( j_0,j_0)$,  $(j_0,j_1)$ and $(j_1,j_0)$ are admissible.
Then the set \begin{equation}\label{Lempty}{\frak L}(A,\mu)\setminus \{0,\pi/2,\pi\}=\emptyset\end{equation} is empty,  which means
\[
L(A,\mu)>0,\qquad  \text{for all}\quad k\in (0,\pi)\setminus \{\pi/2\}.
\]
\end{theorem}

Theorem~\ref{thm3}  provides   an example of a subshift   for which  $ \Omega$ is a proper  subset of  ${\mathcal A}^{\Bbb Z}$,   and yet  the relation   \eqref{Lempty} holds.

A point $p\in \Omega$  is said to be  periodic for $T$ provided  there is  a positive integer $n_p$
for which $T^{n_p}\,p=p$.  The collection of all periodic points of $T$  is denoted in this paper by ${\rm Per}(T)$.

\begin{theorem} \label{thm5} Let $T:\, \Omega\to\Omega$ be  a  subshift of finite type.
Assume that $\mu$ is a $T$-ergodic  measure on $\Omega$ that  has a local product  structure and   the property ${\rm supp}(\mu)=\Omega$.
Let $A$ be defined  by \eqref{defineA}.  Suppose that $T$ has a fixed point in $\Omega$.
Then   
\begin{equation}\notag
  {\frak L}(A,\mu)\setminus\{0,\pi/2,\pi\}=\bigcap_{p\in {\rm Per}(T)}\{k\in (0,\pi/2)\cup (\pi/2,\pi):\,\, k^2\in\sigma(p)\},
\end{equation}
where $\sigma(p)$ denotes   the spectrum of the  Schr\"odinger operator $ H_p$ on the  periodic graph $\Gamma_p$.
\end{theorem}

\section{Proof of Theorem~\ref{thm2}}

As  we mentioned before, a  point $p\in \Omega$  is called  periodic for $T$ provided  there is  a positive integer $n_p$
for which $T^{n_p}\,p=p$.
If  $p\in \Omega $   is  periodic,  then  $A(T^np)$  is a periodic function of $n$,  because 
$A(T^{n_p + n}p) = A(T^np)$ for every $n\in {\Bbb Z}$.  
For a periodic   point  $p$ of period $n_p$, define  $\Delta_p(E)$ to be the trace of  the monodromy matrix $A_{n_p}(p)$ 
$$ \Delta_p(k) ={ \rm Tr}(A_{n_p}(p)).$$
By   ${\rm Per}(T)$,  we denoted  the collection of all periodic points of $T$.

Below,  we   often identify the projective plane $\Bbb{CP}^1$   with   the set ${\Bbb C}\cup \{\infty\}$
meaning that   every vector of the  form $(\xi,1)\in \Bbb{C P}^1$  could be uniquely characterized by  $\xi\in {\Bbb C}\cup \{\infty\}$.
For each $k\in {\Bbb C}_+\cup {\Bbb R}$  such that  $\Delta_p(k)\neq \pm2$, there are exactly two eigendirections $s(k)$ and $u(k)$  in $\Bbb{C P}^1$ of the
monodromy matrix $A_{n_p}(p)$.   In fact,   they are  given by the formulas
\[
s(k)=\frac{a-d+\sqrt{(\Delta_p(k))^2-4}}{2c},\qquad u(k)=\frac{a-d-\sqrt{(\Delta_p(k))^2-4}}{2c}
\]
where $a$  and $b$  are the two elements of  the first  row of the matrix $A_{n_p}(p)$, and $c$ is the first   element of  the second  row.
Since  all solutions $k$ of the  equation  $(\Delta_p(k))^2=4$  are   real,   and all elements of  the matrix $A_{n_p}(p)$ are trigonometric polynomials in $k$,  the functions $s(k)$  and $u(k)$  are  at least meromorphic  in  the open half-plane ${\Bbb C}_+$.
Moreover,  we see that, if $k\in {\Bbb R}$ and  $|\Delta_p(k)|>2$, then $s(k) \neq  u(k)$  are real.   If  $k\in {\Bbb R}$ and $ |\Delta(k)| < 2,$ then $s(k)$ and $u(k)$ are not real.   In the latter case,  we
have $s(k) = \overline{u(k)}$.   

For $\Delta(k)=\pm2$,  the monodromy matrix $A_{n_p}(p)$ either  has a unique
real invariant direction, or it equals $\pm {\rm Id}$.
 We may think of the  first  case as $s(k) = u(k)$.  In the second case, all directions
are invariant.

As  we already mentioned,  the  cocycle $A=A^{(k)}$ is  locally constant.
Put differently,  there is an $\epsilon>0$  such that
$$
A(\omega')=A(\omega)\qquad  \text{whenever}\qquad d(\omega',\omega)<\epsilon.
$$

{\it Definition }.   Let $T:\Omega\to \Omega$ be a subshift of finite type.
The local stable set of a point $\omega\in \Omega$ is defined by
$$
W^s(\omega)=\{\omega'\in\Omega:\,\, \omega'_n=\omega_n\quad \text{for}\quad n\geq 0\}
$$
and the local unstable set of $\omega$  is defined by
$$
W^u(\omega)=\{\omega'\in\Omega:\,\, \omega'_n=\omega_n\quad \text{for}\quad n\leq 0\}.
$$

\bigskip

For $\omega'\in W^s(\omega)$,  define $H_{\omega',\omega}^{s,n}$  to be
$$
H_{\omega,\omega'}^{s,n}=\bigl[A_n(\omega')\bigr]^{-1}A_n(\omega).
$$
Since $d(T^j\omega',T^j\omega)\leq e^{-j}$ tends to $0$ as $j\to \infty$,  there is an index $n_0$  for which
$$
H_{\omega,\omega'}^{s,n}=H_{\omega,\omega'}^{s,n_0}\qquad \text{for }\quad  n\geq n_0.
$$
In this case, we define  the stable  holonomy  $H_{\omega,\omega'}^s$ by
$$
H_{\omega,\omega'}^s=H_{\omega,\omega'}^{s,n_0}.
$$
The unstable holonomy $H_{\omega,\omega'}^u$  for $\omega'\in W^u(\omega)$   is  defined similarly by
$$
H_{\omega,\omega'}^u=
\bigl[A_n(\omega')\bigr]^{-1}A_n(\omega)   \qquad \text{for all}\quad n\leq -n_0.
$$

These abstract definitions  of  holonomies work not only for  the cocycle \eqref{defineA},  but also  for  any locally constant  function  $A:\Omega\to{\rm SL}(2,{\Bbb R})$.
However, if $A$ is   defined by \eqref{defineA}, then the matrices $H_{\omega,\omega'}^s$ and $H_{\omega,\omega'}^u$  become  very specific.

\begin{proposition}
Let $A$ be  defined  in \eqref{defineA}.  Then
\[
H_{\omega,\omega'}^{s}=\bigl[A(\omega')\bigr]^{-1}A(\omega), \qquad  \text{for any}\quad \omega'\in W^s(\omega).
\]
Similarly,
\[
H_{\omega,\omega'}^{s}={\rm Id}, \qquad  \text{for any}\quad \omega'\in W^u(\omega).
\] 
\end{proposition}

The general theory of dynamical systems  tells us that
the  cocycle
$$
(T,A): \Omega\times {\Bbb R \Bbb P}^1\to {\Bbb R \Bbb P}^1
$$
defined  by 
$$
(T, A)(\omega,\xi)= (T\omega, A(\omega) \xi)
$$
has an invariant probability measure $m$  on $\Omega\times {\Bbb R \Bbb P}^1.$ We say that such a measure $m$ projects to $\mu$
if $m(\Delta \times {\Bbb R \Bbb P}^1)=\mu(\Delta)$  for all Borel subsets $\Delta$ of $\Omega$.  Given any $T$-invariant  measure $\mu$  on $\Omega$,
   one can  find a $(T,A)$-ivariant measure  $m$  that projects to $\mu$ by applying  the standard Krylov-Bogolyubov trick used to construct  invariant measures.

\smallskip

{\it Definition}.  Suppose $m$ is  a $(T, A)$-invariant probability measure 
on $\Omega\times {\Bbb R \Bbb P}^1$
that projects to $\mu$.   A disintegration of $m$ 
 is a measurable family $\{m_\omega:\quad \omega\in \Omega\}$ of probability measures on ${\Bbb R \Bbb P}^1$
having the property
$$
m(D)=\int_\Omega m_\omega(\{\xi\in {\Bbb R \Bbb P}^1:\,\, (\omega,\xi)\in D\})d\mu(\omega)
$$
for each measurable set  $D\subset \Omega\times{\Bbb R \Bbb P}^1.$

\smallskip
  Existence of  such a disintegration is guaranteed 
by Rokhlin’s theorem. 
Moreover,
$\{\tilde m_\omega:\quad \omega\in \Omega\}$  is another disintegration of $m$ then  $m_\omega=\tilde m_\omega$  for $\mu$-almost
every $\omega\in \Omega$. 
It is  easy to see  that $m$  is
$(T, A)$-invariant if and only if $A(\omega)_*m_\omega=m_{T\omega}$
for $\mu$-almost every  $\omega\in \Omega$.

\smallskip

{\it Definition}.  
A $(T,A)$-invariant measure $m$  on $\Omega\times {\Bbb R \Bbb P}^1$ that projects to $\mu$ is said to be an su-state for $A$ provided it has 
a disintegration $\{m_\omega:\quad \omega\in \Omega\}$  such that   for  $\mu$-almost every $\omega\in \Omega $, 
$$\,$$
1)$$\quad A(\omega)_* m_\omega=m_{T \omega},$$
2) $$\bigl( H^s_{\omega,\omega'}\bigr)_*m_\omega=m_{\omega'}\qquad
\text{ for every}\quad \omega'\in W^s(\omega).$$
3)$$ \bigl( H^u_{\omega,\omega'}\bigr)_*m_\omega=m_{\omega'}\qquad
\text{ for every}\quad \omega'\in W^u(\omega)$$

\smallskip

The following statement  was  proved in \cite{ADZ}  (Proposition 4.7)  for a   significantly  larger  class of functions $A$.
\begin{proposition} Let $A$ be locally constant.  Suppose $\mu$ has a local product structure   and $L(A,\mu)=0$.   If the support of   the measure  $\mu$  coincides  with all of $\Omega$,  then there exists an su-state for $A$.
\end{proposition}
We apply the  following method to extend $m_\omega$ to a continuous function of $\omega$ on all of $\Omega$.
For each $1\leq j\leq \ell$,  we select     a  point   $\omega^{(j)}$  in $[0;j]\cap \Omega_0$ for which  the measure $m_{\omega^{(j)}}$  is well defined.
Then we set
\begin{equation}\label{definem}
m_{\omega}=\Bigl( H^u_{\omega\wedge \omega^{(\omega_0)}, \omega }   H^s_{\omega^{(\omega_0)}, \omega\wedge \omega^{(\omega_0)}}\Bigr)_*  m_{\omega^{(\omega_0)}}.
\end{equation}
Obviously  $m_\omega$ depends  continuously on $\omega$.

\bigskip

Observe  that ${\Bbb R \Bbb P}^1$ may  be   aslo viewed  as ${\Bbb R}\cup \{ \infty\}$,
because  any vector of the form  $(\xi, 1)\in {\Bbb R \Bbb P}^1$  is uniquily  characterized by $\xi\in {\Bbb R}\cup \{\infty\}$.
Aslo,  ${\Bbb C \Bbb P}^1$ may  be  aslo viewed  as ${\Bbb C}\cup \{ \infty\}$  because  there is a  1:1    mapping of  one set   onto another.
The  part  of  ${\Bbb C \Bbb P}^1$  that is  mapped onto the extended upper half-plane ${\Bbb C}_+\cup \{ \infty\}$   will be denoted  by ${\Bbb C}_+{ \Bbb P}^1$.

Now we  will  state Proposition 4.9  from \cite{ADZ}  in the following more convenient form:
\begin{proposition}\label{barycenter} For each probability measure $\nu$ on ${\Bbb R \Bbb P}^1$
containing
no atom of mass $\geq 1/2$, there is an unique point  $B(\nu)\in {\Bbb C_+ \Bbb  P}$, called the conformal
barycenter of $\nu$, such that 
$$
B(P_* \nu) = P\cdot B(\nu)
$$
 for each $P\in {\rm SL}(2,{\Bbb R})$.
\end{proposition}

Let $m$  be an su-state   with a continuous  disintegration $m_\omega$.
If $m_\omega$   does not have an  atom  of mass $\geq 1/2$, then we set $Z(\omega)\subset {\Bbb C_+ \Bbb  P}$ to   be $\{B(m_\omega)\}$.
Otherwise $Z(\omega)$ is defined to be the   collection of points $\xi $ at  which $m_\omega(\{\xi\})\geq 1/2$.
Since $m_\omega$ is a probability measure,  the set $Z(\omega)$  can  contain at most  two points.
The following  theorem is a consequence of Proposition~\ref{barycenter}.

\begin{theorem} Let $A$ be locally constant.  Suppose $\mu$ has a local product structure   and $L(A,\mu)=0$.   Then
$$
A(\omega) Z(\omega)=Z(T\omega) \qquad  \text{for each}\quad \omega\in \Omega.
$$
 If $\omega',\omega$  are two points in $\Omega$  such that  $\omega_0'=\omega_0$,
then
\begin{equation}\label{Zinvarient}
Z(\omega)=\Bigl( H^u_{\omega\wedge \omega', \omega }   H^s_{\omega', \omega\wedge \omega'}\Bigr) Z(\omega').
\end{equation}
In particular, the number of the points in $Z(\omega)$  does not  depend on $\omega$.   Moreover,  if $Z(\omega)$ is real   for one $\omega$, then it is   real for all $\omega\in \Omega$.
Similarly, if $Z(\omega)$ is not real  for one $\omega$, then it is  not  real for all $\omega\in \Omega$.
\end{theorem}

{\it Proof.} The  last three lines of the theorem follow  from the fact  that    for any two points $\omega$  and $\omega'$ in $\Omega$, there is a  real matrix $P\in {\rm SL}(2,{\Bbb R})$
for which  $Z(\omega)=P\cdot  Z(\omega')$.   Indeed, if $\omega'_0=\omega_0$, then this property is  guaranteed  by  \eqref{Zinvarient}.   On the other hand, since $T$ is transitive,
for any two points $\omega'$ and $\omega$,
there is an index $n$  and a point $\tilde \omega$ such that $(T^n\tilde\omega\bigr)_0= \omega'_0$  while  $\tilde \omega_0=\omega_0$.  Therefore
$$
Z(T^n\tilde \omega)=A_n(\tilde\omega)Z(\tilde\omega)=\Bigl( H^u_{T^n\tilde\omega\wedge \omega', T^n\tilde\omega }   H^s_{\omega', T^n\tilde\omega\wedge \omega'}\Bigr) Z(\omega'),
$$
which implies that
\begin{equation}\label{Ztilde}
Z(\tilde\omega)=[A_n(\tilde\omega)]^{-1}\Bigl( H^u_{T^n\tilde\omega\wedge \omega', T^n\tilde\omega }   H^s_{\omega', T^n\tilde\omega\wedge \omega'}\Bigr) Z(\omega').
\end{equation}
It remains to note that
\begin{equation}\label{Zbez}
Z(\omega)=\Bigl( H^u_{\omega\wedge \tilde \omega, \omega }   H^s_{\tilde\omega, \omega\wedge \tilde\omega}\Bigr) Z(\tilde\omega).
\end{equation}
$\Box$

\bigskip

\begin{corollary}\label{corH} Let $A$ be defined  by \eqref{defineA}.  Suppose $\mu$ has a local product structure. 
Let \[{\frak L}(A,\mu)=\{k\in [0,\pi]:\quad 
 L(A,\mu)=0\}. \]  Then  for each pair  of points $\omega$  and $\omega'$ in $\Omega$,  there is an analytic  function  ${\frak H}_{\omega,\omega'}:  {\Bbb C}\to {\rm SL}(2,{\Bbb C})$  
for which
$$
{\frak H}_{\omega,\omega'}(k)Z(\omega')=Z(\omega) \qquad  \text{for all }\quad k\in {\frak L}(A,\mu),
$$
and   ${\frak H}_{\omega,\omega'}(k)\in {\rm SL}(2,{\Bbb R})$  for all $k\in [0,\pi]$.
\end{corollary}

{\it Proof}. This  statement is a   consequence of  the  relations \eqref{Ztilde} and \eqref{Zbez}. $\,\,\,\,\,\,\Box$

\bigskip

\begin{proposition}
Let $A=A^{(k)}$ be  the cocycle \eqref{defineA} and let $k\in (0,\pi)\setminus \{\pi/2\}$.  Let $\omega$  be a fixed  point of $T$.
Assume that  $L(k)=L(A,\mu)=0$.
Then  $Z(\omega)$   consists of  one point  $e^{ik}$ and  is not real.
\end{proposition}

{\it Proof.}   If $\omega$ is a fixed  pont, then $\omega_n=\omega_{n-1}$  for all $n$.
Therefore,
$Z(\omega)$ is invariant with respect to the  linear transformation
\[
 A(\omega)=
 A^{(k)}(\omega)= \begin{pmatrix}
2\cos(k)& -1\\
1& 0
\end{pmatrix}
\]
This matrix has   two distict  invariant  directions  $(e^{\pm ik},1)$, which  implies that $Z(\omega)=\{e^{ik}\}$. $\,\,\,\,\,\Box$

\section{End  of the proof of Theorem~\ref{thm2}}

If $p$ is  a  periodic  point  of the mapping $T$, then  by the symbol $n_p$,   we  denote  the smallest period of $p$.  

\begin{proposition}  \label{infmanyK} Let $A$ be defined  by \eqref{defineA}.  Suppose $\mu$ has a local product structure. 
 Let $p$  be a fixed  point of $T$, and let $q$ be another periodic point of $T$.
Assume that   the set  \[{\frak L}(A,\mu)=\{k\in [0,\pi]:\quad 
 L(A,\mu)=0\}\] 
contains infinitely many points. Then  there is an eigendirection $e(k)$ of the matrix $A_{n_q}(q)$  such that
\begin{equation}\label{He=e}
{\frak H}_{q,p}(k)e^{ik}=e(k) \qquad  \text{for all }\quad k\in {[0,\pi]},
\end{equation}
where ${\frak H}$ is the  same as in Corollary~\ref{corH}.
 \end{proposition}

{\it Proof.}  Without loss of generality, we may assume that  there is a closed bounded  interval $I\subset [0,\pi]$ and a  converging sequence $k_j\in ({\rm Int}\,I)\cap {\frak L}(A,\mu)$,
such that
\[
|\Delta_q(k)|<2\qquad \forall k\in {\rm Int}\,I,
\]
and all elements  of   the sequence $k_j$  are distinct.  For each $k\in I$,  let $e(k)$  be the  eigendirection of $A_{n_q}(q)$  that belongs to ${\Bbb C_{+}}{ \Bbb P}$ (the upper half-plane).
Then
$$
{\frak H}_{q,p}(k_j)e^{ik_j}=e(k_j) \qquad  \text{for all }\quad j\in {\Bbb N}.
$$
Consequently, \eqref{He=e}  holds  by analyticity of the functions appearing on different sides. $\,\,\,\,\,\Box$

\bigskip

For     a  periodic  point  $p\in \Omega$,  we consider the periodic operator $\tilde H_p$  defined on $\ell^2({\Bbb Z})$  by
\begin{equation}\label{opHtilde}
\bigl[\tilde H_p u\bigr](n)=\frac{p_n}{p_n+p_{n-1}}u(n+1)+\frac{p_{n-1}}{p_n+p_{n-1}}u(n-1),\qquad   \forall u\in \ell^2({\Bbb Z}).
\end{equation}
The spectrum of  this operator is  the union of finitely many closed intervals called "bands"  separated  by  finitely many gaps.
The following  statement that is  well known  consequence of the equation \eqref{SEquation}.

\begin{proposition}
Let  $p$ be   a  periodic  point  of the mapping $T$.  Let $k\in [0,\pi]$.  If  the eigendirections of $A_{n_p}(p)$    are not real,  then $\cos(k)$
belongs to one of the  bands of the spectrum of the periodic operator $\tilde H_p$  defined in \eqref{opHtilde}.
\end{proposition}

Combining this  proposition with  Theorem~\ref{infmanyK}, we obtain   the  followng result.
\begin{corollary}\label{antiKorot}
Let $A$ be defined  by \eqref{defineA}.  Suppose $\mu$ has a local product structure. 
Assume that $T$ has a fixed  point,  and
 that   the set  \[{\frak L}(A,\mu)=\{k\in [0,\pi]:\quad 
 L(A,\mu)=0\}\] 
consists of  infinitely many points. Then for any periodic point $p\in \Omega$,  the spectrum  of $\tilde H_p$  coincides  with $[-1,1]$.
\end{corollary}

The statement below is a  consequence of Theorem 1.2 of the paper \cite{Korotyaev}.

\begin{theorem}\label{Korot}
Let  $p$ be a  periodic  point  of the mapping $T$.  If $n_p>1$,   then  the spectrum of $\tilde H_p$ has at least one  open gap
contained in $[-1,1]$.
\end{theorem}

We see that the  conclusion of Corollary~\ref{antiKorot} contradicts Theorem~\ref{Korot}. That means  the  assumptions of  Corollary~\ref{antiKorot}  cannot be fulfilled.
Thus, the set ${\frak L}(A,\mu)$ is finite.  $\,\,\,\,\Box$

\section{Proof of Theorem~\ref{thm3}}

Clearly,
Theorem~\ref{thm3}  provides   an example of a subshift   for which  $ \Omega$ is a proper  subset of  ${\mathcal A}^{\Bbb Z}$,   and yet  $${\frak L}(A,\mu)\setminus \{0,\pi/2,\pi\}=\emptyset.$$

We argue by contradiction.  Assume that $A$ is defined  by \eqref{defineA}  and  $L(A,\mu)=0$ for some $k\in[0,\pi]\setminus \{0,\pi/2,\pi\}$.
  Observe that  $A$ depends  only on   the  two coordinates  $\omega_{-1}$  and $\omega_0$   of $\omega$.
Therefore,   all   unstable holonomies are    identity operators, while stable holonomies are  the  matrices $[A(\omega')]^{-1}A(\omega)$.  Consequently,  if $m_\omega$  is a continuous  disintegration of  an  su-state,  then
$$
m_{\omega}=m_{\omega'},\quad \text{whenever}\quad  \omega_{-1}= \omega_{-1}' \quad \text{and}\quad \omega_0=\omega'_0.
$$
But then  the equality  
$$
A(\omega)m_\omega=m_{T\omega}
$$  implies 
that $m_{T\omega'}=m_{T\omega}$  whenever $ \omega_{-1}=\omega'_{-1}$ and $ \omega_0=\omega'_0$.    

This property can be formulated  in terms of the cylinder sets, of the form
\[
[n; j_0, j_1] = \{\omega\in\Omega:\quad
\omega_{n+i}= j_i, \quad 0 \leq  i \leq 1\}
\]
with $n\in {\Bbb Z}$ and  $j_0, j_1 \in {\mathcal A}$.  Namely,   for each pair $j_0,j_1$ of  symbols  in $\mathcal A$, the function  $\omega\to m_\omega$ is constant on the sets 
\[
[-1; j_{0}, j_1] \quad {\rm and}\quad [-2; j_{0}, j_{1}],
\]
provided that these sets are not empty.

Let us now  give   a  condition  that  makes  the latter property  impossible.
First,  define ${\mathcal W}_2$ as the set  of words $(j,j')\in{\mathcal A}\times {\mathcal A}$  of length two  that are allowed in $\Omega$:
\[
{\mathcal W}_2=\{(j,j')\in{\mathcal A}\times {\mathcal A}: \exists \omega\in \Omega \,\, \text{such that}\,\, \omega_{-1}=j,\,\omega_0=j'\}.
\]
Observe that if $(j_{-2},j_{-1})\in{\mathcal W}_2$  and $(j_{-1},j_0)\in{\mathcal W}_2$,  then
\[
  m_\omega=m_{\omega'}\qquad \forall \omega\in [-1; j_{-1},  j_0],\,\omega'\in [-2; j_{-2},j_{-1}].
\]
The latter  follows  from   the fact  that there is at least one   $\tilde \omega \in [-1; j_{-1}  j_0]\cap [-2; j_{-2},j_{-1}]$
and the  property  that the  function $\omega\to m_\omega$  is constant on each of these cylinders.
Consequently,   $m_\omega$  depends  only on $\omega_{-1}$   in the sense that
\[
  m_\omega=m_{\omega'}\qquad \text{whenever}\quad \omega_{-1}=\omega'_{-1}.
\]

Suppose that  there are two distinct letters $j_0,j_1$ in ${\mathcal A}$ such that the words   $( j_0,j_0)$,  $(j_0,j_1)$ and $(j_1,j_0)$ belong to  ${\mathcal W}_2$.
Then
\[
q=\dots,j_0,j_0,j_0,j_0,j_0,j_0,j_0,j_0,j_0,j_0,j_0,j_0,\dots \quad \text{ is a  fixed  point of }\,T,
\]
while
\[
p=\dots,j_0,j_1,j_0,j_1,j_0,j_1,j_0,j_1,j_0,j_1,j_0,j_1,\dots \quad \text{ is a  periodic  point of }\, T.
\]
Since $q_0=p_0$, we conclude that
\[
m_p=m_q.
\]
Therefore,
\begin{equation}\label{Zp=Zq}
Z(p)=Z(q).
\end{equation}
One the other  hand,  the relation
\[
A^{(k)}(\omega)\cdot Z(\omega)=Z(T\omega), \qquad \forall \omega\in \Omega.
\]
leads to the  equality 
$$
A_2(p)\cdot Z(p)=Z(T^2p)=Z(p).
$$ 
Combining this relation with \eqref{Zp=Zq}, we  obtain that, if $k\neq \pi/2$, then
\begin{equation}\label{A2=e}
A_2(p)\cdot e^{ik}=e^{ik}.
\end{equation}
 Let us now show  that this   cannot be true.
It follows  from \eqref{defineA}  that
\[
 A_2(p)=\sqrt{\frac{p_{-1}}{p_{0}}} \begin{pmatrix}
\frac{p_{-1}+p_{0}}{p_{-1}}\cos(k)& -\frac{p_{0}}{p_{-1}}\\
1& 0 \end{pmatrix}
\sqrt{\frac{p_0}{p_{-1}}}
 \begin{pmatrix}
\frac{p_0+p_{-1}}{p_0}\cos(k)& -\frac{p_{-1}}{p_0}\\
1& 0
\end{pmatrix}=
\]
\[
 \begin{pmatrix}
(x+2+1/x)\cos^2(k)-x&- (1+1/x)\cos (k)\\
(1+1/x)\cos (k) & -1/x
\end{pmatrix},\qquad \text{where}\quad x=\frac{p_0}{p_{-1}}.
\]
Consequently, \eqref{A2=e} can be written in the form
\[
\bigl((x+2+1/x)\cos^2(k)-x\bigr)e^{ik}- (1+1/x)\cos (k)= e^{ik}\bigl(
(1+1/x)\cos (k) e^{ik} -1/x\bigr).
\]
This means  that $z=e^{ik}$ is a root   of the quadratic equation
\[
(1+1/x)\cos (k) z^2 -\Bigl(1/x+\bigl(x+2+1/x)\cos^2(k)-x\bigr)\Bigr)z+(1+1/x)\cos (k)=0.
\]
In particular,  the real part of the root $z=e^{ik}$  equals
\[
{\rm Re}\, z=\cos(k)=
\frac{ 1/x+\bigl(x+2+1/x)\cos^2(k)-x\bigr)}{2(1+1/x)\cos (k)},
\]
which implies that $x=1/x$.  Thus,  $p_0=j_0=p_{1}=j_1$.
The obtained contradiction shows that our assumption was incorrect and $L(A,\mu)>0$.  $\,\,\,\,\,\,\Box$

\section{Proof of Theorem~\ref{thm5}}

 As we mentioned before, if $q$  is a fixed point of $T$, then $Z(q)=\{e^{ik}\}$  for all $k\in  {\frak L}(A,\mu)\setminus\{0,\pi/2,\pi\}$.
Therefore, $Z(\omega)$ is not real for all $k\in  {\frak L}(A,\mu)\setminus\{0,\pi/2,\pi\}$  and all $\omega\in \Omega.$
In particular, $Z(p)$ is not real for all $p\in {\rm Per}(T)$  and $k\in  {\frak L}(A,\mu)\setminus\{0,\pi/2,\pi\}$.
Consequently,   $A_{n_p}(p)$  has a  complex  eigenvalue, which implies that $ k^2\in\sigma(p)$.
Thus, 
\begin{equation}\notag
 {\frak L}(A,\mu)\setminus\{0,\pi/2,\pi\}\,\subset \bigcap_{p\in {\rm Per}(T)}\{k\in (0,\pi)\setminus\{\pi/2\}:\,\, k^2\in\sigma(p)\}.
\end{equation}
Conversely,   let $k\in(0,\pi)\setminus\{\pi/2\}$   satisfy the condition  $k^2\in \bigcap_{p\in {\rm Per}(T)} \sigma(p)$.  We must show that $L(A^{(k)},\mu)=0$.
For this purpose, we  set
$$
L(A,p)=\lim_{n\to\infty}\frac1n \ln\bigl(\|A_n(p)\|\bigr),\qquad \forall  p\in {\rm Per}(T).
$$
If  $p$ is a periodic point and $k^2\in \sigma(p)$, then
\begin{equation}\label{L=0}
L( A,p)=0.
\end{equation}
Now we  use  the following  result  proved in a much more general setting by Kalinin (see  Theorem 1.4  in \cite{K}).
\begin{proposition}\label{approxL} Let $A$  be defined by \eqref{defineA}.
 Then  for each $\delta>0$ there is a periodic point $p\in  \Omega$  such that $|L(A,p)-L( A, \mu)|<\delta$.
\end{proposition}
Combining Proposition~\ref{approxL}  with the  equality \eqref{L=0}, we obtain that $$L( A, \mu)=0.$$
Thus, $k\in {\frak L}( A,\mu)$.  
$\,\,\,\,\,\Box$

\end{document}